\documentclass[aps,pre,twocolumn,superscriptaddress,,amsmath,amssymb,]{revtex4-1} % for review and submission  %% superscriptaddress,
\usepackage{graphicx}% Include figure files
\usepackage{xcolor}
\usepackage{dcolumn}% Align table columns on decimal point
\usepackage{bm}% bold math
\usepackage[normalem]{ulem}
\usepackage[utf8]{inputenc}
\usepackage[T1]{fontenc}
\usepackage{mathptmx}
\usepackage{etoolbox}
\usepackage[normalem]{ulem}
\usepackage{amsmath}
\begin{document}
\title{Absorption of electromagnetic waves at oblique resonance   in  plasmas  threaded by inhomogenous magnetic fields}

\author{Trishul Dhalia}
\email{trishuldhalia@gmail.com}
\author{Rohit Juneja}
\author{Amita Das}
\email{amita@iitd.ac.in}
\affiliation{Department of Physics, Indian Institute of Technology Delhi, Hauz Khas, New Delhi 110016, India
}

\begin{abstract}

There has been significant interest lately in the study of Electromagnetic (EM) waves interacting with magnetized plasmas. The variety of resonances and the existence of several pass and stop bands in the dispersion curve for different orientations of the magnetic field offer new mechanisms of EM wave energy absorption \cite{PhysRevE.105.055209,Juneja_2023,vashistha2020new}. However, earlier studies have  investigated only special cases of magnetized plasma geometry (e.g., RL mode $(\vec{k}||\vec{B}_{ext}$) or $(\vec{k}\perp\vec{B}_{ext})$ X,O-mode configuration). In these specific cases, EM waves encounter specific resonances (e.g. for $(\theta=0)$  cyclotron resonances,  and for $(\theta=\pi/2)$,  hybrid resonances). A general  case of  EM wave propagation is at an oblique angle with respect to the externally applied magnetic field $\vec{B}_{ext}$ has been considered here.  Furthermore, the magnetic field is chosen to be inhomogeneous such that the EM wave pulse encounters a resonance layer within the plasma medium. A 2-D Particle-In-Cell (PIC) simulation using the OSIRIS 4.0 platform has been carried out for these studies. A significant enhancement in absorption leading to almost complete absorption of laser energy by the plasma has been observed. A detailed study characterizing the role of the external magnetic field profile, EM wave intensity, etc., has also been carried out.
\end{abstract}
\keywords{Electromagnetic wave plasma interaction, resonance, magnetized plasma modes, laser energy absorption}
\maketitle
\section{Introduction}
 Collision-less electromagnetic (EM) wave energy absorption in plasmas has attracted interest due to a variety of applications in various contexts. For instance, applications like fusion, ion acceleration, plasma heating, and direct laser acceleration (DLA) \citep{kaw2017nonlinear, das2020laser,nishida1987high,ganguli2016development,ganguli2019evaluation,gibbon1996short,pukhov1999particle,khan2023enhanced}  employ EM wave energy absorption to achieve their objectives. Thus theoreticians and experimentalists are always looking for new methods of enhancing the energy absorption of EM waves in plasmas. 
   The known methods in the context of un-magnetized plasmas rely on collisional schemes \cite{braginskii1958transport,balescu1960irreversible,van2022collisional}. Non-collisional mechanisms like resonant, vacuum heating, $\vec{J} \times \vec{B}$ processes, etc. are also employed, especially at high energies where the collisional transfer turns out to be inefficient.  The resonant absorption occurs when the frequency of the EM wave matches the local plasma frequency and part of incident EM wave energy is transferred to electrostatic oscillations \cite{kruer1988physics}.   %Hence the absorption is confined only to the resonant layer in an inhomogeneous plasma. 
   The other two mechanisms (vacuum and $\vec{J} \times \vec{B}$) rely on the randomization of electron trajectories as they get pulled out and pushed inside the plasma at the vacuum plasma interface \cite{PhysRevLett.59.52,kruer1988physics}. The mode conversion from EM to electrostatic plasma waves at resonance has been extensively studied theoretically \cite{gershman1962propagation} and demonstrated experimentally by \cite{kim1974development} in the context of microwaves interacting with plasma. 
  The availability of high-power sources (e.g. up to 10  PW now for lasers) makes nonlinear phenomena inevitable.  Features like high harmonic generation, parametric instability, wakefield generation, wave-breaking, etc., are routinely observed in the context of laser-plasma interaction. Some of these nonlinear processes often influence the process of energy absorption. The exploration of the role of nonlinear effects in the context of energy absorption for microwave plasma interaction is also of importance.
 
 Newer ways and means to enhance energy absorption of EM waves in plasmas have always been of paramount interest. Magnetized plasma offers a larger variety of normal modes compared to the unmagnetized case. It also permits the propagation of EM waves inside the overdense plasma medium. The laser-plasma interaction studies, however, have remained restricted to the un-magnetized regime. The laser frequency is very high, even eliciting a magnetized response from a lighter electron species in this case requires a magnetic field of the order of $100's$ of kilo tesla. This was out of the realm untill recently. Recent technological advancements have produced magnetic fields in the range of kilo tesla in the laboratory \cite{nakamura2018record} and there are proposals for attaining fields as high as mega tesla level \cite{korneev2015gigagauss}. This development,  therefore, also opens up the possibility of exploring the  laser interactions with magnetized plasmas.

\begin{figure*}
  \centering
  \includegraphics[width=16cm]{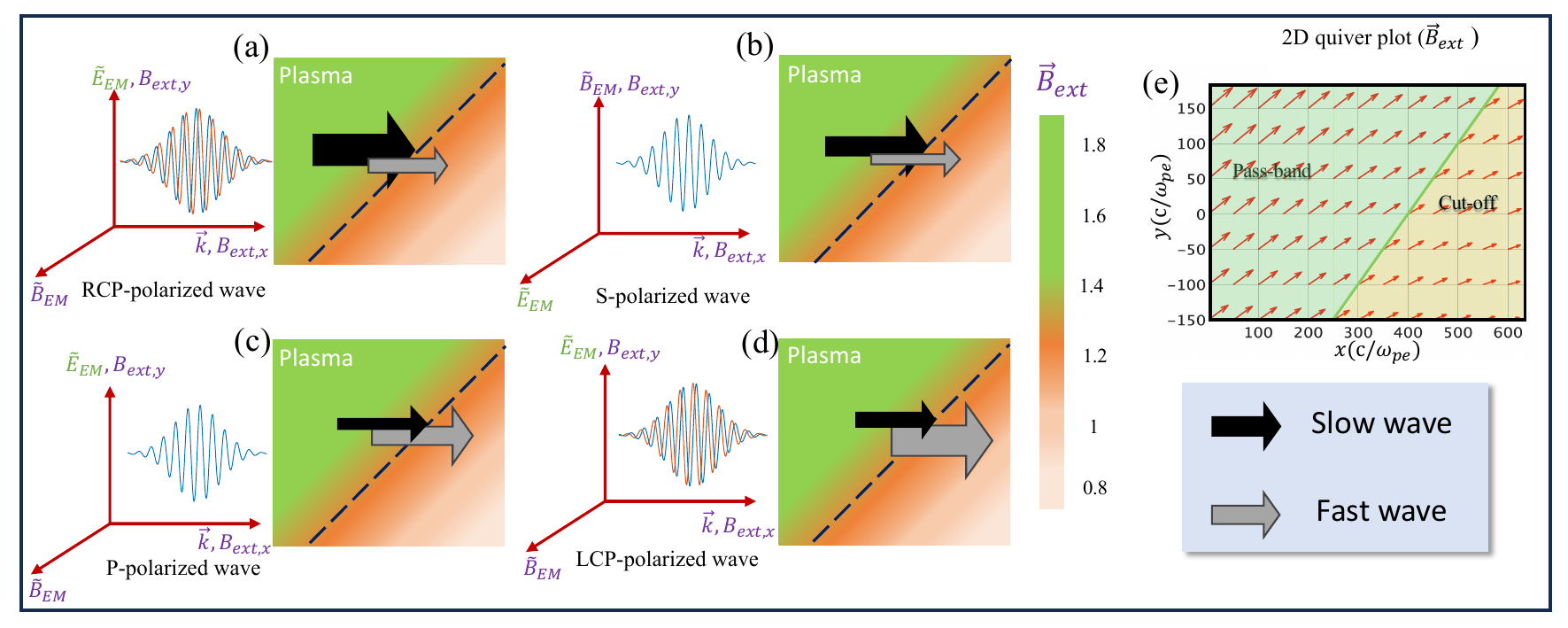}% Images in 100% size
  \caption{The Figure here demonstrates the schematic representation (not to scale) of the geometry chosen for simulation. We have carried out a 2D particle-in-cell simulation of an EM wave interaction with a rectangular plasma slab of homogeneous density. A spatially varying external magnetic field $(\vec{B}_{ext})$ of oblique orientation has been applied as shown in figure (e). The incoming EM wave breaks into slow and fast waves. The slow wave stops near resonance layer and the fast wave moves out of the plasma undamped. Figures $(a,b,c,d)$ represent EMF energies going into slow and fast wave for each case of polarization. The sizes of the arrows represent relative strength in each wave. }  
\label{fig:schematic}
\end{figure*}

Microwave frequencies being low, their interactions with plasmas threaded with external magnetic fields often elicit a magnetized response from plasmas, for which the gyrofrequency of one or both species can be made higher than the microwave radiation frequency rather easily. For the case of microwaves, nonlinear absorption of free electron laser $(<1\textrm{GW})$ generated intense microwave pulses near electron cyclotron resonance have been measured at MTX tokamak \cite{nevins1987nonlinear,allen1994nonlinear}. The magnetic field in tokamaks, however, is fairly complicated with a sheared geometry and inhomogeneity.

%\textcolor{red}{  Recently, pulsed high-powered Microwave sources of $4.6 GW$ and frequency $9.96GHz$ have been achieved in experiments using relativistic backward wave oscillators \citep{xiao2016axial,xiao2020efficient}. This recent development of pulsed high-powered microwaves thus opens up an uncharted territory of nonlinear physics in the context of microwave plasma interaction in the magnetized plasma domain. }
 
Thus, the regime of EM wave interactions with magnetized plasmas has attracted significant interest lately, both for lasers as well as microwave fields. Theoretical and simulation studies carried out recently in this direction have already spelled out several interesting effects.  New mechanisms of laser energy absorption, higher harmonic generation, parametric processes, etc., have been observed in these simulations \citep{das2020laser,vashistha2021excitation,mandal2020spontaneous,mandal2021electromagnetic,kumar2019excitation,goswami2022observations,maity2021harmonic,dhalia2023harmonic,juneja2024enhanced}. 

Most of the studies conducted recently in the context of lasers have been confined to very idealized and specific geometries. The externally applied magnetic field has mostly been  chosen as constant  and its orientation is either along the $X$ or  $R-L$ mode configuration \cite{Juneja_2023,vashistha2020new,vashistha2021excitation,maity2021harmonic,dhalia2023harmonic}, which correspond to the applied magnetic field either along the oscillating magnetic field direction or along the propagation direction, respectively \cite{chen1984introduction}. We investigate the interaction of EM waves in a generalized scenario where the magnetic field is oriented at an arbitrary angle relative to the direction of EM wave propagation. This is done through 2-D particle-in-cell (PIC) simulations using the OSIRIS4.0 platform. Furthermore, the  applied magnetic field is chosen to be inhomogeneous so that the laser encounters the resonance layer within the bulk plasma medium.  We observe that such a choice provides the possibility of greater efficiency in the process of energy absorption.  The condition for this kind of resonance to form and energy absorption to occur have been investigated analytically and by simulation in detail in this manuscript.

 The paper is organized as follows: Section \ref{sec:simulation} describes the simulation geometry and the choice of parameters. Section \ref{sec:theory} describes the theoretical analysis associated with EM wave interactions with plasmas under the obliquely applied external magnetic field. In section \ref{sec:Energy-abs} we carry out a comparative study of the energy absorption of EM waves in the presence  of the resonance layer by considering various polarizations. In section \ref{sec:dual-resonance}, other  characteristics of the  resonance are studied. For instance, the influence of the spatial profile of the magnetic field, the underlying nonlinear features, and parametric processes are discussed. The intensity dependence on the efficiency of energy absorption is also studied. In section \ref{sec:summary}, we summarize our findings.

\begin{table*}
  \begin{center}
\def~{\hphantom{0}}
  \begin{tabular} {|c|c|c|c|}
  \hline
      $$\textbf{ Parameters}$$       & $$ \textbf{ Normalized values}$$   &   $$  \textbf{ Microwave System}$$ & $$  \textbf{Laser System}$$  \\[3pt]
      \hline
    Frequency($\omega_{EM}$)  & 1.2$\omega_{pe} $ & $4.2\times 10^{10}$ rad$s^{-1}$ & $0.2\times 10^{15}$ rad$s^{-1}$ \\ 
    \hline
    Wavelength($\lambda_{EM}$) & 5.23$c/\omega_{pe}$ &  44.88 mm  &$ 9.42 \mu$m \\
    \hline
    Intensity ($I_0$)           & $a_0=0.08$        &  $4.3\times 10^{10} W\textrm{m}^{-2}$ & $1.33\times 10^{19} W\textrm{m}^{-2}$\\
    \hline
                         \multicolumn{4}{|c|}{\textbf{Plasma Parameters}} \\
    \hline
      Density$(n_{e,i})$   & 1  & $3.85\times 10^{17}\textrm{m}^{-3}$ &  $4.47\times 10^{25}\textrm{m}^{-3}$\\
      \hline
     Electron Plasma frequency($\omega_{pe})$   & 1 & $3.5\times10^{10}$ rad$s^{-1}$ & $3.77\times 10^{14} $ rad$s^{-1}$ \\
     \hline
      Electron skin depth($c/\omega_{pe})$ & 1 & 8.57 mm & $0.79 \mu$m\\
      \hline
  \end{tabular}
  \caption{Simulation parameters are shown here in normalized as well as in corresponding SI units}
  \label{tab:parameters}
  \end{center}
\end{table*}

\section{Simulation Details}\label{sec:simulation}
The simulation geometry is depicted in figure \ref{fig:schematic}, and the corresponding simulation parameters are provided in table \ref{tab:parameters}. The laser is incident normal to the plasma slab of constant density as it propagates along the $x$ axis. The plasma is underdense ($\omega_{EM}>\omega_{pe}$). The externally applied magnetic field lies in the $x-y$ plane and is given by the expression: 
\begin{multline}
     \vec{B}_{ext}=[(1.5-0.001x+0.001y)\hat{i}+\\
     (0.863-0.001x+0.001y)\hat{j}]B_N
     \label{eq:bext}
\end{multline}
Here $B_N$ is the normalizing magnetic field described later. The external magnetic field is chosen to be obliquely oriented and linearly decreasing from a high magnetic field regime ($\omega_{ce} > \omega_{EM}$) to a low magnetic field regime ($\omega_{ce} < \omega_{EM}$), ensuring that an electron cyclotron resonance condition lies within the plasma. The chosen magnetic field profile also preserves the condition $\vec{\nabla} \cdot \vec{B} = 0$.  The choice of various polarizations of the incident electromagnetic wave pulse, along with the resonance layer that they would encounter in the plasma for the above 
choice of the external magnetic field has been shown in the four subplots (a) to (d) of the figure. The origin of this resonance layer has analytically discussed in next section. A fully relativistic, massively parallel PIC code, OSIRIS 4.0 \citep{hemker2000particle, fonseca2002osiris, fonseca2008one}, has been used for simulation. OSIRIS uses normalized parameters, where the time and length scales are normalized by $t_N\rightarrow \omega_{pe}^{-1}$ and $l_N\rightarrow c/\omega_{pe}$, respectively. The simulation box has a longitudinal extent of $600l_N$ and, along the transverse direction, runs from $-150l_N$ to $150l_N$. Here $l_N$ denotes the plasma skin depth $c/\omega_{pe}$, where $c$ and $\omega_{pe}$ represent the speed of light and electron plasma frequency.  We have a plasma boundary that starts from $100l_N$ to $500l_N$ along the longitudinal direction, while in the transverse direction, it extends from $-130$ to $130 l_N$. For this study, completely ionized electron-ion plasma is considered. Where the ion-electron mass ratio is 1837, the grid sizes are taken to be $dx,dy=0.1c/\omega_{pe}$. Both electron and ion are mobile in the simulation box. The time step for the simulation is set to $dt=0.02\omega_{pe}^{-1}$. The number of macro-particles is chosen as $4\times 4=16$ per cell. The incident EM  pulse profile has been chosen to be of polynomial form in the longitudinal direction while it has a Gaussian profile along the transverse direction. The EM wave enters from the wall on the left side, with a pulse duration of $150 \omega_{pe}^{-1}$ and spot size of $100c/ \omega_{pe}$. Traveling in a forward direction it will  focus at $x=100l_N$ on the vacuum plasma interface. The electric and magnetic fields are normalized by $E_N = B_N = m_ec\omega_{pe}/e$ where $m_e$ and $e$ are electron mass and charge, respectively. The choice of normalized parameters has been shown in  table \ref{tab:parameters}. The corresponding values for the microwave as well as the laser system have also been depicted. The microwave system features a pulsed, high-power microwave with a frequency of $6.7$ GHz and power of approximately $4$ GW, directed at a large plasma box measuring approximately $341$ cm in length and $85.3$ cm in diameter. The necessary external magnetic field ranges from about $0.14$ T to $0.34$ T \cite{singh1992experimental,mahaffey1963microwave}. The laser system utilizes a high-intensity PW, $\textrm{CO}_2$ laser with a wavelength of $9.42\mu$m, targeting an underdense plasma with dimensions of $0.31$ mm in thickness and $0.2$ mm in width. The required magnetic field strength for this setup ranges from about $1.6$ kT to $3.7$ kT \cite{nakamura2018record}. While these parameters are not derived from specific experiments, they are well-suited to inspire new experiments in this field.

\begin{figure}
    \centering
    \includegraphics[width=8cm]{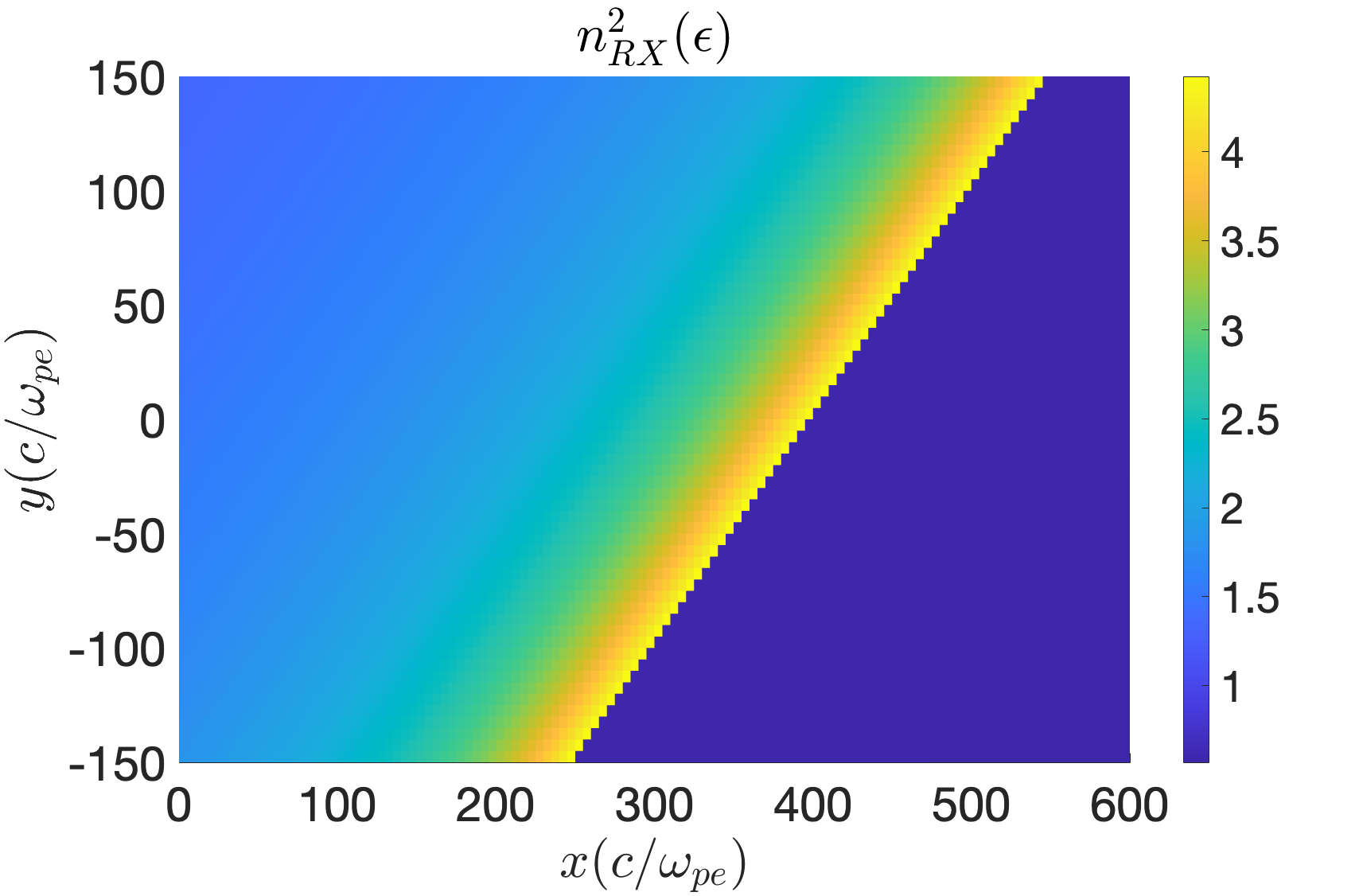}
    \caption{Variation in the refractive index $n_{RX}^2$ across space for a slow wave. The cut-off line corresponds to the resonance layer ($\omega_{ce}=\omega_{EM}$) at $y=x-395$.}
    \label{fig:ref_index}
\end{figure}

\begin{figure*}
    \centering
    \includegraphics[width=14cm]{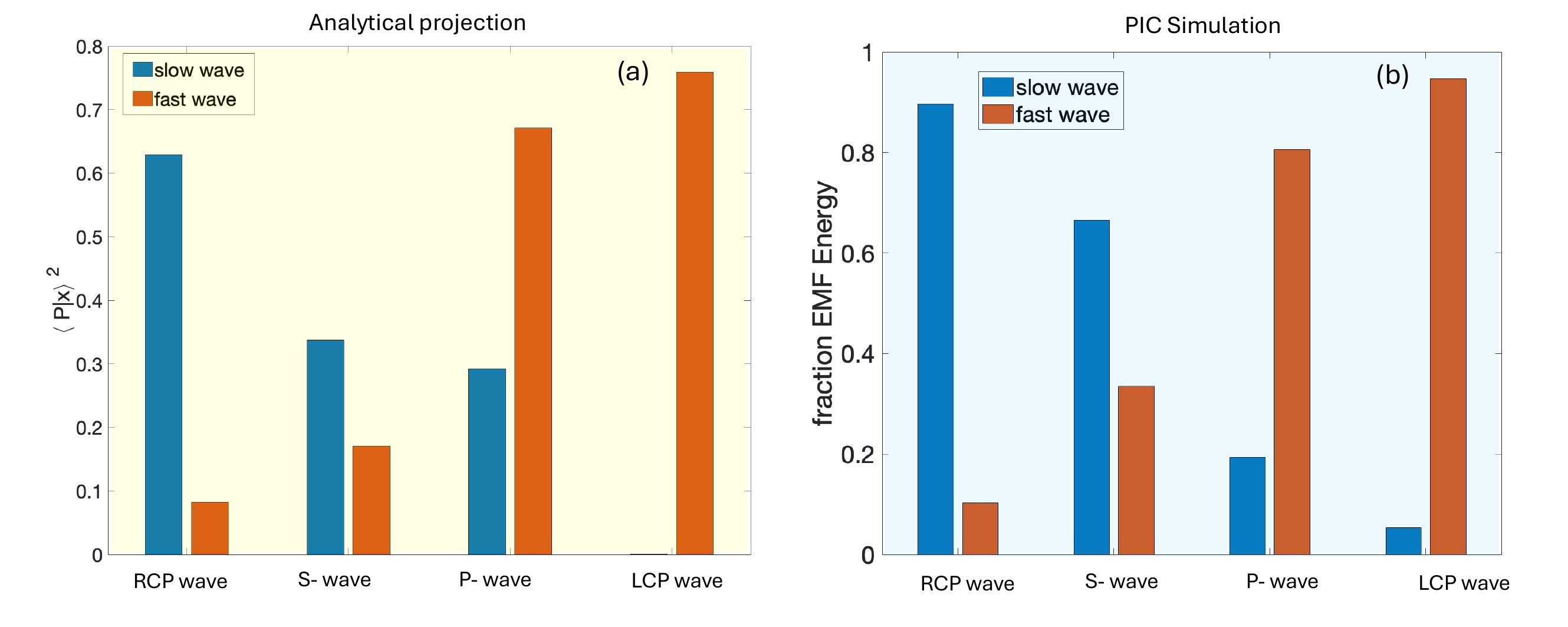}
    \caption{Figure (a) depicts the projection of each type of polarized wave onto the slow and fast waves at the plasma surface, as derived from analytical calculations. Figure (b) represents the fraction of the electromagnetic field energy that is transmitted into the slow and fast waves within the plasma. }
    \label{fig:bar_chart}
\end{figure*}

\begin{figure*}
    \centering    \includegraphics[width=12cm]{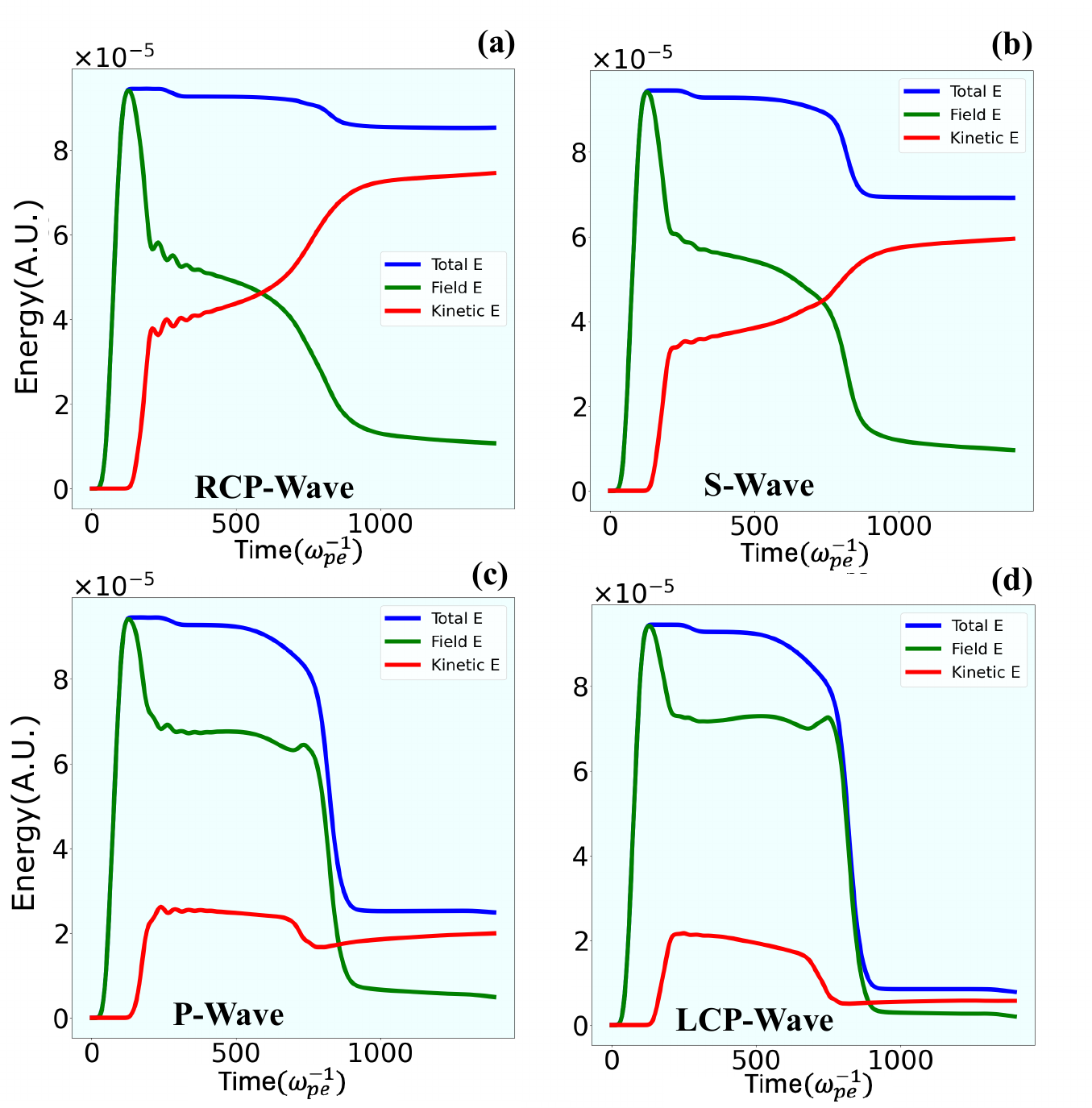}
    \caption{Figure demonstrates time evolution of total energy, field energy, and kinetic energy of electrons in the simulation box.  }
    \label{fig:energy_evolution}
\end{figure*}

\section{Theoretical Analysis}\label{sec:theory}
 The most general dispersion relation for  an electromagnetic wave  of frequency $(\omega_{EM})$ in a  magnetized cold plasma  propagating at some angle $\theta$ with respect to applied external magnetic field $\vec{B}_{ext}$ is given by \cite{stix1992waves}, \\
\[
\begin{pmatrix}
    \textrm{S}-n^2\cos^2\theta & -i\textrm{D}  & n^2\cos^2\theta \sin^2\theta \\
    i\textrm{D}       &     \textrm{S}-n^2    &         0               \\
    n^2\cos^2\theta \sin^2\theta  &  0  &   \textrm{P}-n^2\sin^2\theta 
\end{pmatrix} 
\begin{pmatrix}
    E_x \\ 
    E_y\\ 
    E_z
\end{pmatrix} =
\begin{pmatrix}
    0 \\
    0\\ 
    0 
\end{pmatrix}
\] \\
To achieve a nontrivial solution, the determinant of the above matrix has to be zero, which gives us the following dispersion relation:
\begin{equation}
    \tan^2\theta =\frac{\textrm{P}(n^2-\textrm{R})(n^2-\textrm{L})}{\textrm{(S}n^2-\textrm{RL})(n^2-\textrm{P})}
    \label{dispersion}
\end{equation}
In our simulations, the electron response time scales, $(\omega_{ce},\omega_{pe})$, are much faster than the ion response time scales, $(\omega_{ci},\omega_{pi})$. Therefore, ions primarily serve as a neutralizing background, while the electron responses dictate the interaction processes at these faster 
time scales. For the electron species, these coefficients are given as, $\textrm{P}=1-\frac{\omega_{pe}^2}{\omega_{EM}^2}$, $\textrm{S}= 1-\frac{\omega_{pe}^2}{\omega_{ce}^2-\omega_{EM}^2}$  ,$\textrm{D}=- \frac{\omega_{ce}\omega_{pe}^2}{\omega_{EM}(\omega_{ce}^2-\omega_{EM}^2)}$, R$=$ S$+$D and L$=$ S$-$D.
For our geometry, we have chosen propagation direction of the incoming EM wave along $x-$direction and external magnetic field  is chosen to lie  in the $x-y$ plane as per the equation (\ref{eq:bext}), 
Thus,  the angle $\theta$ is a function of space, 
\begin{equation}
    \tan\theta(x,y)= \frac{0.863-0.001x+0.001y}{1.5-0.001x+0.001y}
    \label{theta}
\end{equation}
We have considered completely ionized homogeneous plasma and normalized plasma frequency is $\omega_{pe}=1$, and EM wave frequency is $\omega_{EM}=1.2\omega_{pe}$. The electron cyclotron frequency varies in space as  $\omega_{ce}(x,y)=e|\vec{B}_{ext}(x,y)|/mc$. Solving equation  (\ref{dispersion}) using (\ref{theta}) for $n^2$  provides two solutions. One corresponds to the slow wave and the other is the fast wave, characterized on the basis of their group velocity. The CMA diagram dictates that for our geometery in underdense plasmas, the slow wave lies in the R-X wave normal surface, whereas the fast wave lies in the L-O wave normal surface \cite{stix1992waves,swanson2020plasma}. Thus, we chose to refer to the slow wave as the R-X wave and the fast wave as the L-O wave. The fast L-O wave does not stop inside the plasma and  travels out of the plasma unhindered.    The refractive index $n_{RX}^2$ for the slow R-X wave has plotted in figure \ref{fig:ref_index}. The refractive index rises as the electromagnetic wave travels within the plasma, and near $y=x-395$ line, $n_{RX}^2$ shows a discontinuous layer. At this oblique layer inside the plasma, EM waves resonate with the electrons, and localized absorption occurs near this region.

We consider four different polarizations of the electromagnetic wave. They can be represented by the electric field vector as follows: 
\begin{equation}
  \tilde{E}_{EM}= E_{EM,0}(\alpha \hat{y}+\beta\hat{z} )e^{i(\vec{k}.\vec{r}-\omega t)}   
\end{equation} 
For RCP (right-hand circular polarization) and LCP (left-hand circular polarization), we have $\alpha = 1$ and $\beta = \pm i$ respectively. The S-polarization corresponds to $\alpha = 0$ and $\beta = 1$(the electric field of EM wave oriented outside the 2-D plane $(x-y)$ considered for simulation),  for P-polarization (electric field of EM wave oriented along the 2-D simulation plane) for which we have 
$ \alpha =1 $ and $\beta = 0$. It should be noted that the external magnetic field $(\vec{B}_{ext})$ (\ref{eq:bext}) has both $x$ and $y$ components which vary in space. 
Each of these  four cases of incident polarizations  get divided into slow and  fast  waves inside the plasma. The fraction of energy that gets converted into slow and fast waves is observed to  depend on the initial choice of the polarization of the EM wave.

\begin{figure*}
  \centering
  \includegraphics[width=12cm]{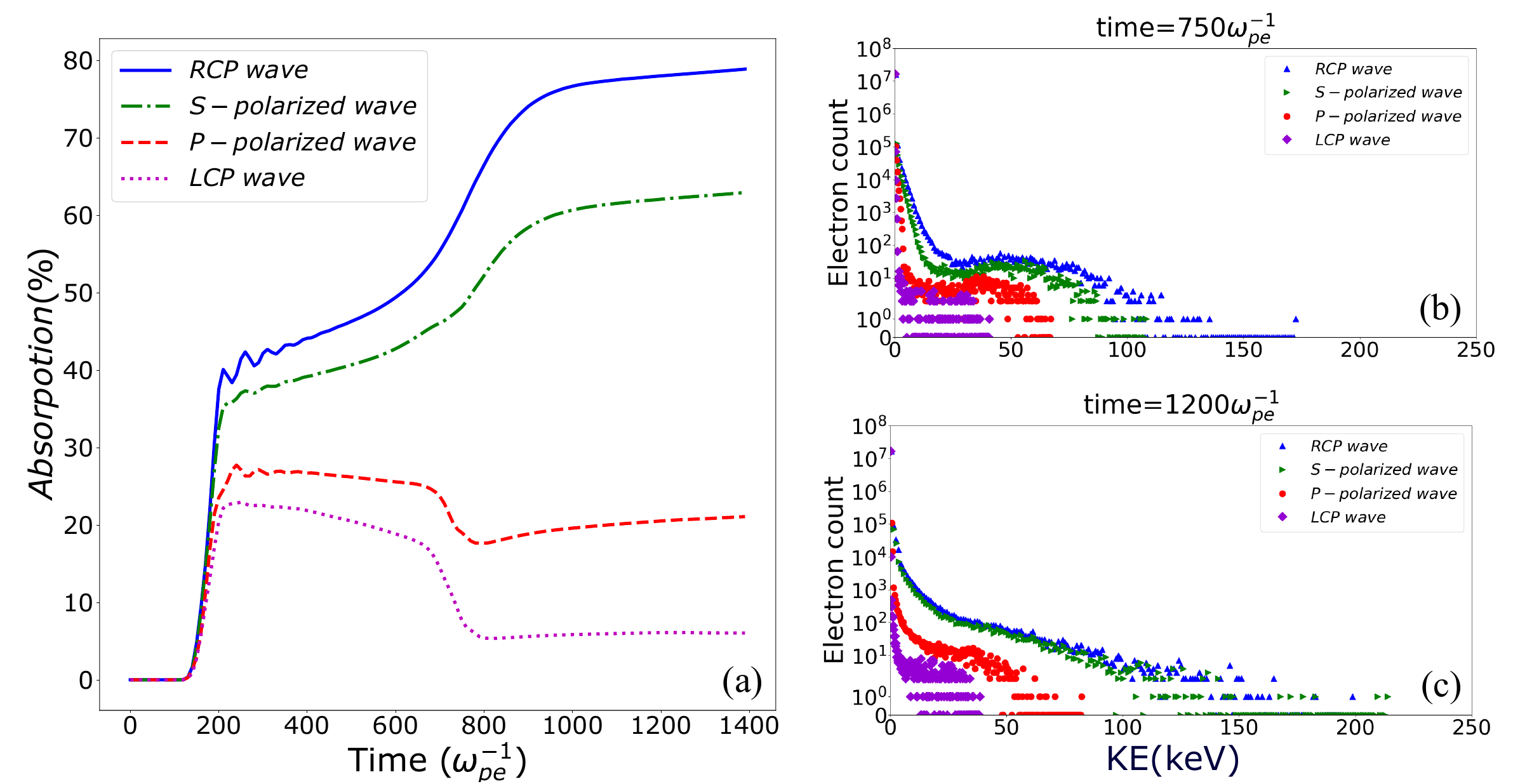}% Images in 100% size
  \caption{Figure $(a)$ illustrates the absorption $\%$ of EM waves inside the magnetized plasma slab in time for all four cases. Figure $(b,c)$ represents the distribution of kinetic energy of electrons at times $750, 1200\omega_{pe}^{-1}$ respectively for these different waves. }
\label{fig:energy_with_pol}
\end{figure*}

The projection of the incident wave into the slow and the fast wave can also be calculated based on the response matrix at the vacuum plasma interface. By plugging the value of $n^2$ for the slow wave ($n^2_{RX}$) and the fast wave ($n^2_{LO}$) into the matrix, four independent eigenvectors are possible. Two of the eigenvectors will correspond to reflected  waves in vacuum, while the other two will represent transmitted slow ($|P_{\textrm{slow}}\rangle$) and transmitted fast wave ($|P_{\textrm{fast}}\rangle$) in the plasma. 
% eigemodese Since determinant of the above matrix is set to 0. Which constrains the $3\times 3$ matrix to give only two eigenmodes corresponding to each wave. These two eigenmodes will represent transmitted and reflected wave at the plasma vacuum surface. We are interested in the transmitted slow and fast wave corresponding to incident wave. Thus, for each polarization, we project the transmitted eigenvector corresponding to slow and fast wave onto the incident wave.
In matrix representation, the incident polarized waves can be modeled in normalized form as, 
\[|X_{RCP}\rangle=\frac{1}{\sqrt{2}}\begin{pmatrix}
    0 \\
    1 \\
    i 
\end{pmatrix}, 
|X_{S}\rangle=\begin{pmatrix}
    0 \\
    0 \\
    1 
\end{pmatrix},\]
\[|X_{P}\rangle=\begin{pmatrix}
    0 \\
    1 \\
    0 
\end{pmatrix},
|X_{LCP}\rangle=\frac{1}{\sqrt{2}}\begin{pmatrix}
    0 \\
    1 \\
    -i 
\end{pmatrix}
\]
We calculated the fraction of energy of these incident waves that gets converted into transmitted slow and fast waves by taking the square of the inner product with the eigenvectors $|P_{\textrm{slow}}\rangle$ and  $|P_{\textrm{fast}}\rangle$  at the vacuum plasma interface. Figure \ref{fig:bar_chart} (a) shows a bar chart of the projection of different polarizations into slow and fast wave. These analytical calculations show an exquisite resemblance to the fraction of electromagnetic field energy of the incident wave pulse transmitted into the slow and fast waves in plasma, as demonstrated by the PIC simulation shown in Figure \ref{fig:bar_chart}(b).

In the next section, we study the characteristic behavior of the oblique resonance layer and the factors influencing the process of energy absorption. 

\begin{figure*}
  \centering
  \includegraphics[width=14cm]{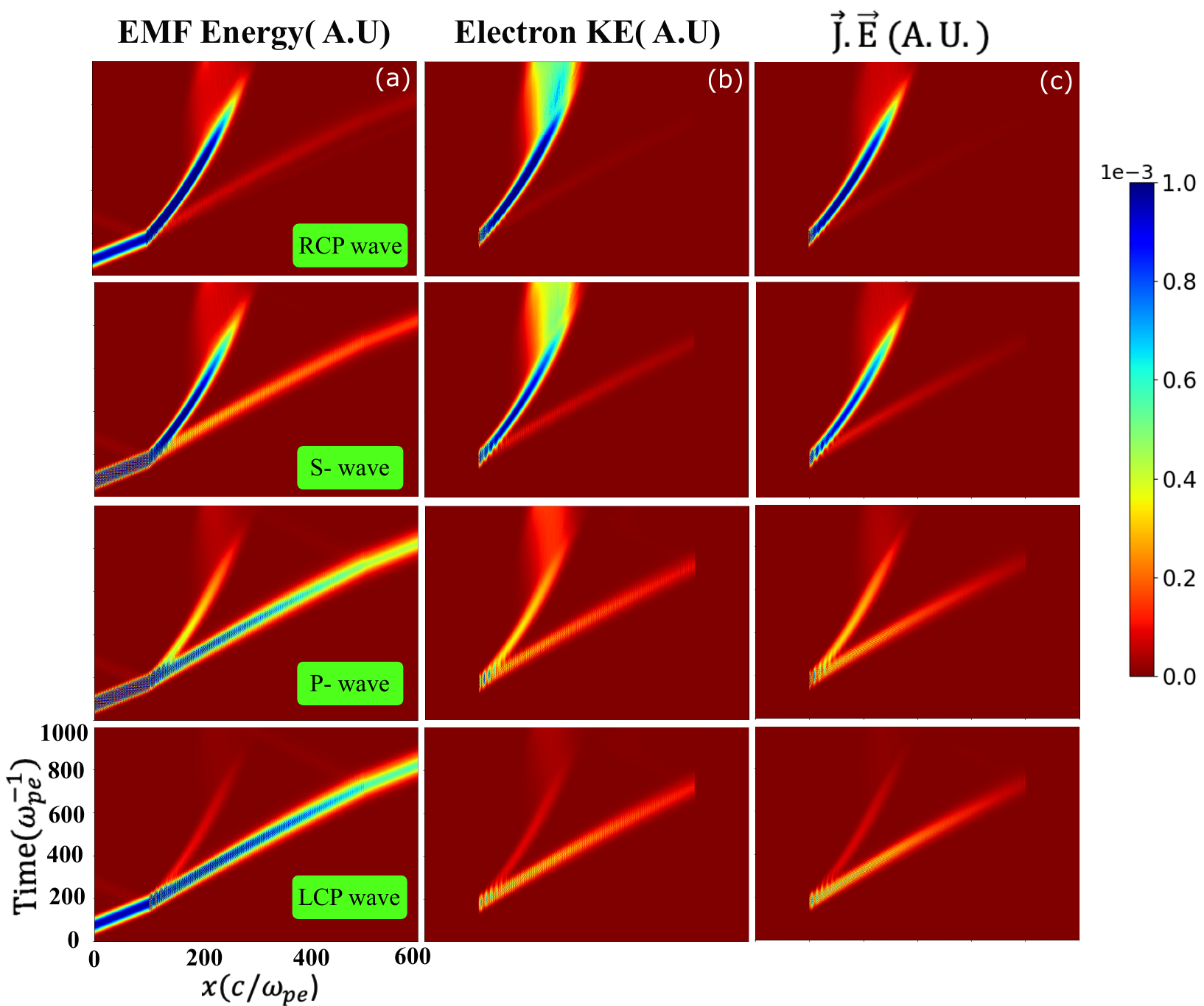} \textbf{}% Images in 100% size
  \caption{ The space variation in x-direction of (a) EMF energy, (b) electron kinetic energy and (c) $\vec{J}\cdot\vec{E}$  averaged along the y-direction, has been plotted against time for each of the four cases in their respective rows. }
\label{fig:localized_absorption}
\end{figure*}
\begin{figure}
    \centering
    \includegraphics[width=6cm]{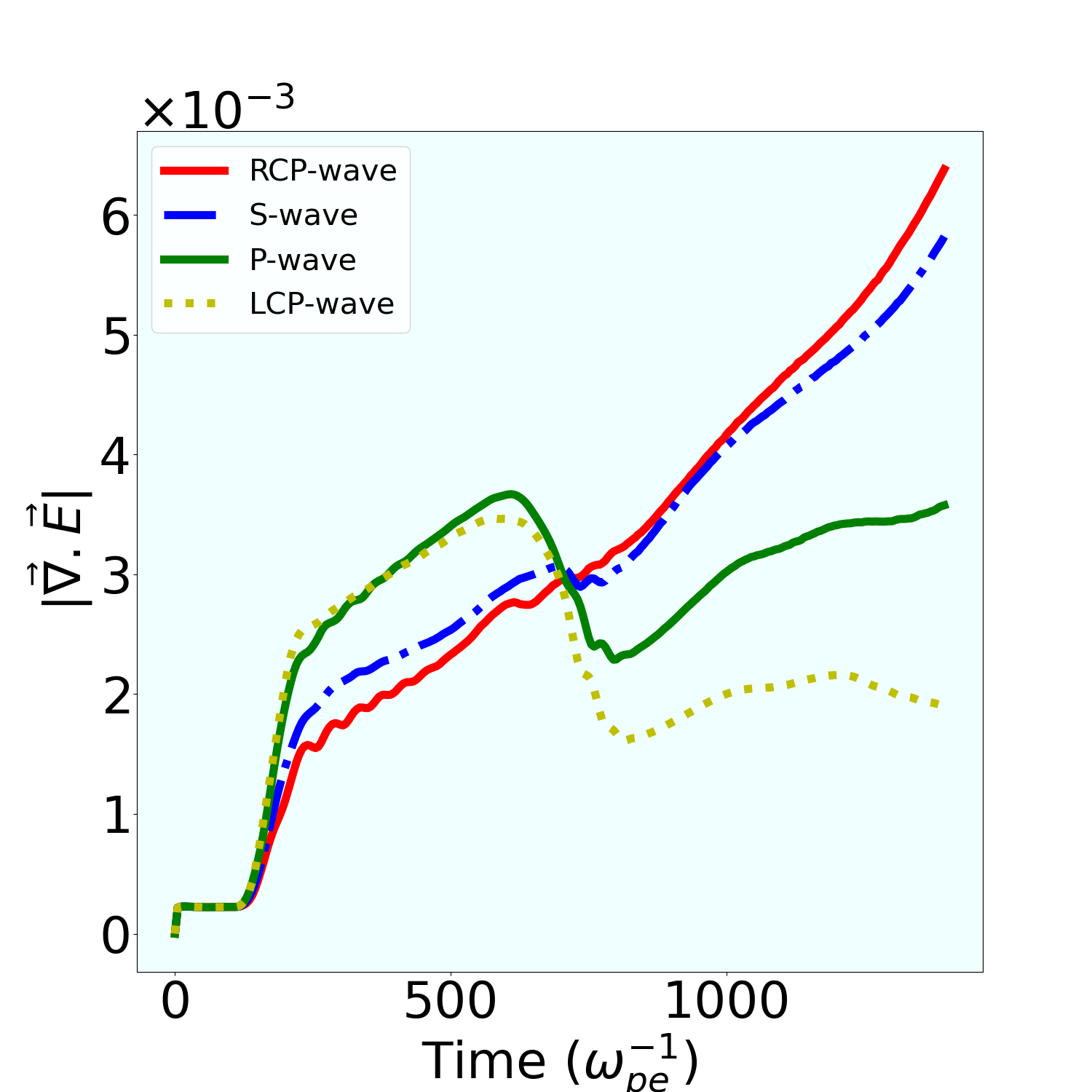}
    \caption{Figure illustrates the time evolution of $\mid\vec{\nabla}.\vec{E}\mid$ for each case of polarized wave.}
    \label{fig:dive}
\end{figure} 

\begin{figure*}
  \centering
  \includegraphics[width=12cm]{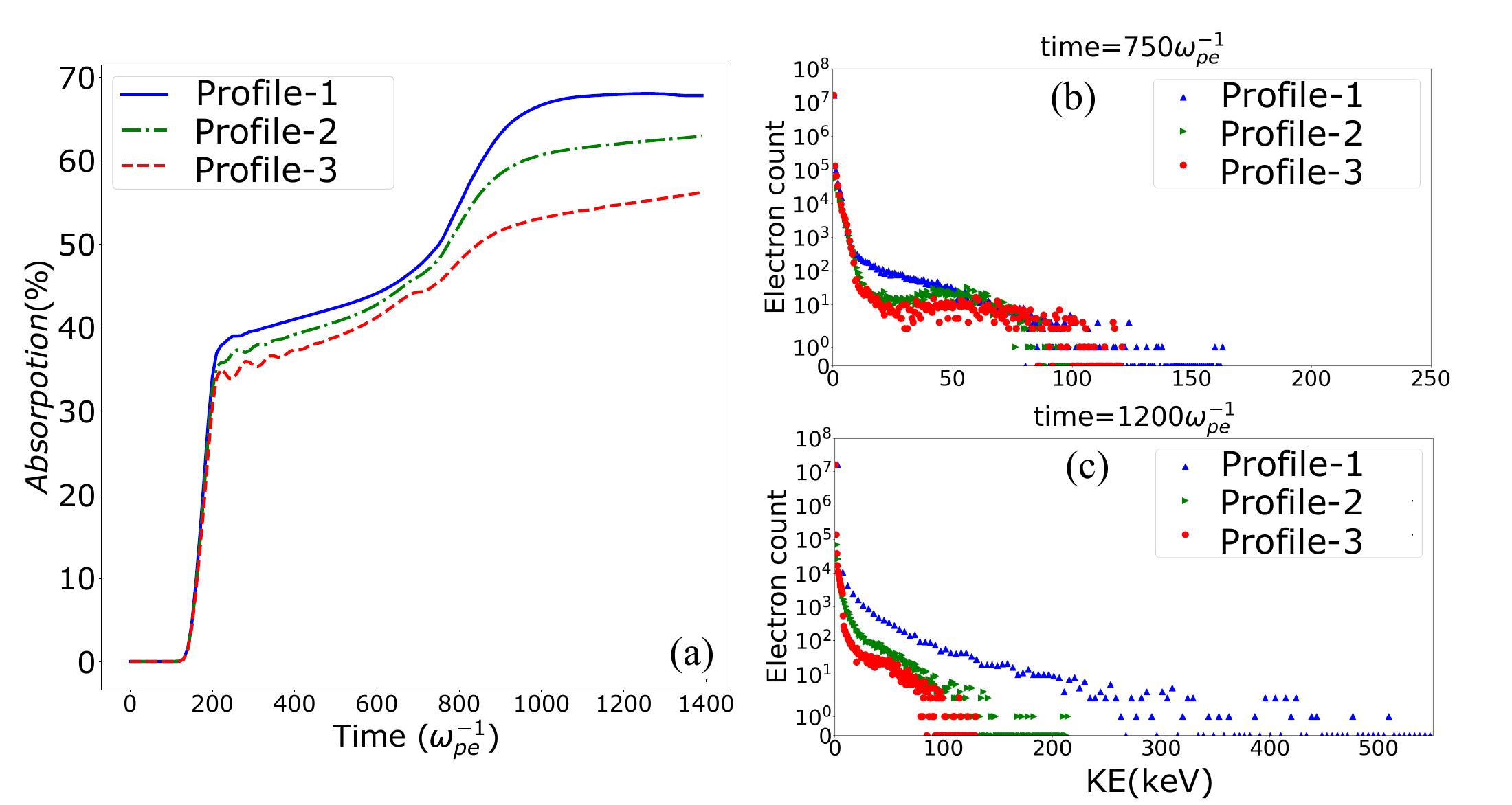}% Images in 100% size
  \caption{Absorption $\%$ of an S-polarized wave has been plotted in figure $(a)$ for the three different magnetic field geometries as a function of different external magnetic field profiles. figures $(b,c)$ show distribution of electron kinetic energy at times $750$ and $1200\omega_{pe}^{-1}$ respectively.}
\label{fig:energy_with_distance}
\end{figure*}

\begin{figure*}
    \includegraphics[width=12cm]{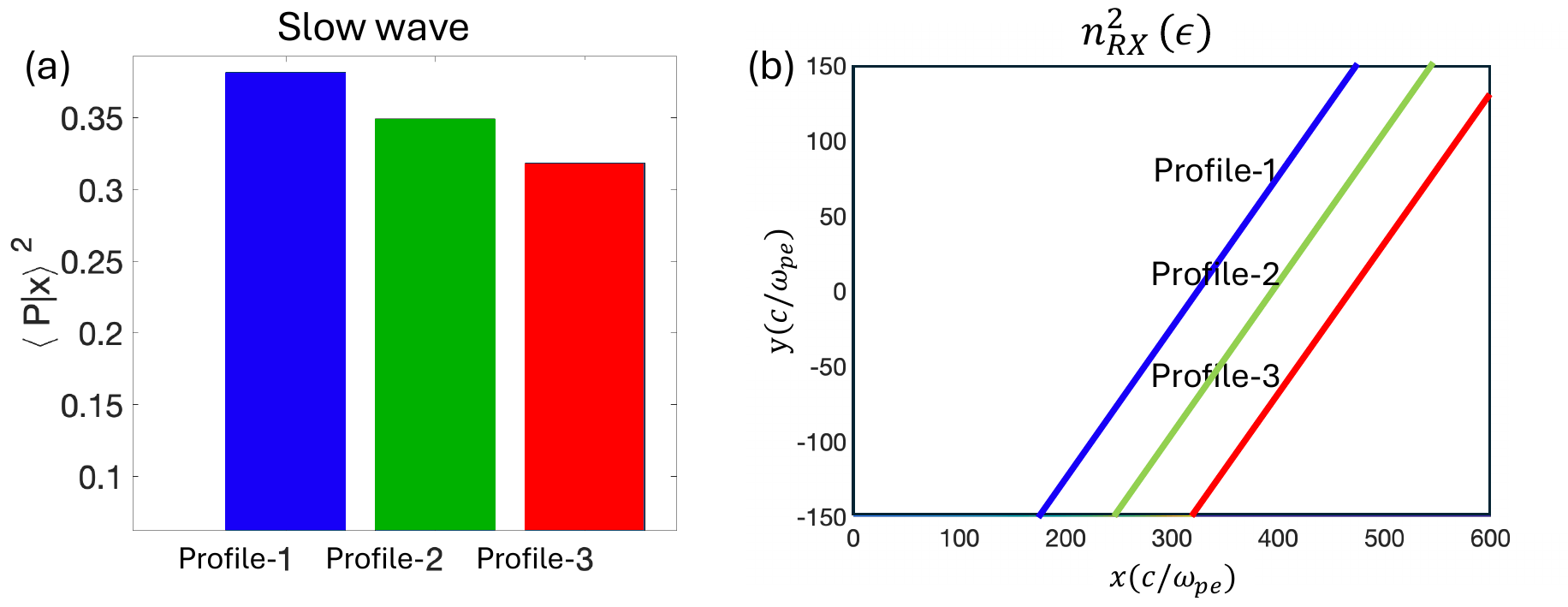}
    \caption{ Figure (a) illustrates the fraction of electromagnetic field (EMF) energy converted to the transmitted slow wave at the vacuum-plasma interface for three different external magnetic field profiles. Figure (b) displays the variation in the refractive index, $n_{RX}^2$, of the slow wave across space in the three cases.}
    \label{fig:ref_index_shift}
\end{figure*}

\section{Energy Absorption}\label{sec:Energy-abs}
 Figure \ref{fig:energy_evolution} depicts the evolution of the total energy, the electromagnetic field energy, and the kinetic energy of electrons for the four different cases of polarization that we have considered. The two dips in the total energy essentially occur when the EM wave pulse leaves the simulation box. The first dip corresponds to a small reflected wave leaving the simulation box whereas the second one occurs when the propagating EM pulse leaves the simulation box from the other side. It should be noted that we are considering an underdense plasma. 
In all four subplots of the figure, wherever the electromagnetic energy dips there is a concomitant increase in the value of electron kinetic energy. This shows that there is an energy transfer from the field energy to the electron kinetic energy. 
%Since the parameters have been tailored so that the EM wave encounters only electron-related resonance the energy transfer occurs only in the electron species here. 
 
 This comparison between the four chosen polarized cases in figure \ref{fig:energy_evolution} elucidates that for RCP and S-polarized waves, field energy gradually decreases in time while kinetic energy surges. This indicates a direct transfer of EM field energy into kinetic energy of the electrons. On the other hand, for P-polarized and LCP waves, only a very small portion is converted into kinetic energy of electrons.
 %\textcolor{red}{It is also worth noting that the RCP and the S wave encounter dual resonance and the energy transfer process from the field energy to the particle kinetic energy is high.  Whereas for the LCP wave and the P wave which only encounter one resonance, and most of the energy leaves the simulation box.} 
The $\%$ of energy absorption (as kinetic energy of the electrons) for the four different cases are compared in figure \ref{fig:energy_with_pol}$(a)$. The plot clearly shows that for the cases of RCP and S waves, the absorption is high. The absorption percentage in the case of RCP is as high as about  
$80\%$ However, for LCP and P-polarization, the absorption efficiency is quite low. 

One can see that there are about four distinct phases of energy evolution.  These phases occur as the EM wave is found to split into fast and slow-moving components. The first phase starts a little early at $t = 200\omega_{pe}^{-1}$ when the EM wave hits the plasma boundary. There is a steep rise in the energy absorption here as the electrons quiver with the incident EM wave field. After this, there is a slow phase of increased energy absorption untill about $t = 650\omega_{pe}^{-1}$ or so for RCP and S -polarised waves. 
 On the other hand, the percentage of absorption gradually decreases in the case of P-polarized and LCP waves during this interval.  At around $t= 650 \omega_{pe}^{-1}$ the fast wave component crosses the plasma boundary at $x=500 c/\omega_{pe}$. In case of, the RCP and S-polarized waves, the fast wave component carries a smaller fraction of the energy compared to the slow wave component. Conversely, for P-polarized and LCP waves, the fast wave component carries a larger fraction of the energy. Thus, when the fast wave leaves the box, the kinetic energy of P-polarized and LCP cases registers some decrease, whereas  no such reduction is observed for RCP and S-polarized waves. On the other hand, one observes a steady increase in the kinetic energy. This can be understood by realizing that the slow waves which carry significant energy in this case at this time, inches closer towards the resonance layer and it's energy gets transferred to electron kinetic energy. This is also evident from the $x-t$ plots in figure \ref{fig:localized_absorption} discussed later. This shows that, in the presence of an oblique resonance layer, polarization plays a major role in energy absorption. Analytical calculations of the absorption coefficient for oblique and perpendicular propagation, as demonstrated by \cite{bornatici1982theory}, show that it is strongly dependent on polarization. Our PIC simulation results and analytical findings are consistent with those reported in \cite{bornatici1982theory}.
 %Thereafter, only the slow wave component is left in the simulation plasma box. This slow wave is responsible for the second phase of slope change in the energy absorption plot. 

 We have shown the electron energy distribution for all four cases at  $t=750 \omega_{pe}^{-1}$ in figure \ref{fig:energy_with_pol}$(b)$. One can observe two clear humps in the distribution function, indicating a two-temperature behavior that the distribution function acquires. This distribution will, therefore, be susceptible  Bump-on-tail instability. The cut-off of the electron kinetic energy is also higher for RCP and S-polarized incident waves, as expected. The maximum cut-off energy of electrons is $175$ keV and $110$ keV for the RCP and S-polarized EM waves respectively, at this time. At a later time after $t=1200 \omega_{pe}^{-1}$ when the $\%$ absorption for all four cases nearly saturates, $80\%$ absorption is achieved for the RCP wave, $65\%$ for the S-polarized wave, $22\%$ for the P-polarized wave and just $7\%$ absorption for LCP wave. Figure \ref{fig:energy_with_pol}$(c)$ shows the distribution of electrons at later time 1200 $\omega_{pe}^{-1}$. The bump appears to have almost vanished, indicating diffusion in velocity space which is like  thermalization  \cite{10.1063/1.860809}. This can also be viewed as the  quasilinear flattening of the bump on the tail \cite{ochs2022momentum, Diamond_Itoh_Itoh_2010}. The maximum cut-off energy of electrons reaches about $220$ keV for RCP waves. For LCP and P-polarized waves, the maximum cut-off is a meager $40$ keV and $90$ keV, respectively. 

The splitting of the EM wave into fast and slow waves can be observed from 
figure (\ref{fig:localized_absorption}). We chose four different polarizations for the incident wave, and the first column of the subplots in this figure shows how the EM wave energy changed over time as a function of the $x$ coordinate, while the $y$ dependence has been averaged over. In all these instances, the EM wave sub-divides into fast and slow components. With time represented by the y coordinate, the slow wave exhibits a steeper slope, and the fast wave shows a larger slope. It can be observed that for the 
RCP and the S-polarized waves most of the wave energy goes into the slow wave component. The electron kinetic energy shown in the middle column maximize at the location of the slow wave component. The plots of the $y$ averaged $\vec{J}\cdot \vec{E}$ indicating the energy transfer process is shown in the last column. The intensity is high at the same space-time location where the electron kinetic energy is found to be high. Unlike RCP and S-polarized waves we observe that for the LCP and P-polarized waves, the fast wave component carries the majority portion of energy. However, figure \ref{fig:localized_absorption} clearly shows that this energy does not get absorbed and ultimately is carried away by the wave from the right-hand side of the plasma boundary.

It was shown in one of our earlier works \citep{Juneja_2023} that when the EM wave encounters a cyclotron resonance there is a direct transfer of EM wave energy to the kinetic energy of the species. However, at plasma wave resonance and hybrid resonances  (lower hybrid and upper hybrid), it is always mediated by an electrostatic field generation. We have also plotted evolution of the average of $\mid \vec{\nabla} \cdot \vec{E} \mid$ in figure (\ref{fig:dive}) for all four cases to understand the generation of electrostatic fluctuations in the plasma medium as these waves propagate. It should be noted that the value of $\mid \vec{\nabla} \cdot \vec{E} \mid$ shows a very rapid rise initially for both the LCP and the P-polarized waves. However, this is only the case while the fast EM wave component remains in the simulation box. It primarily appears to be linked to the reversible energy exchange between the fast EM wave field and the oscillating velocity of the electrons. As soon as this component leaves the simulation box the value of  $\mid \vec{\nabla} \cdot \vec{E} \mid$ diminishes. This time also corresponds to the electron kinetic energy dip in the energy plot of the figure \ref{fig:energy_evolution} for both LCP and the P-polarized waves. On the other hand, the RCP wave and S-polarized waves which encounter the oblique resonance layer because of the slow R-X wave, the increase in $\mid \vec{\nabla} \cdot \vec{E} \mid$ is comparatively at a slower pace, but this continues until the very end. Although the energy absorption is higher from the very beginning in this case, it appears that the direct resonance transfer, not mediated by the generation of an electrostatic field, plays a significant role.

\section{Other characteristic features}\label{sec:dual-resonance}
In this section, we discuss some other characteristic features of the absorption process that have been observed in the context of the present study with the oblique external magnetic field. In particular, we address three distinct features in the three subsections below. The first concerns the influence of the shift in location of the resonance layer and its impact on energy absorption.  The second relates to the underlying nonlinear and phase mixing processes which in general occur along with the absorption of EM wave energy. Lastly, the impact of nonlinear processes on absorption has also studied.  

\subsection{Effect of magnetic field profile}\label{sec:shift_res_points}
Our simulations have demonstrated that the appearance of an oblique resonance layer in the case of RCP and S-polarized waves leads to better absorption. Here, we investigate the possibility of  shifting the position of the resonance layer to see how energy absorption changes. For this purpose, we consider the   S-polarized incident EM wave.

We consider two other profiles of the external magnetic field for this purpose as given below.
\begin{multline}
     \vec{B}_{ext}=[(1.4-0.001x+0.001y)\hat{i}+\\
     (0.863-0.001x+0.001y)\hat{j}]B_N
     \label{profile-1}
\end{multline}
\begin{multline}
     \vec{B}_{ext}=[(1.6-0.001x+0.001y)\hat{i}+\\
     (0.863-0.001x+0.001y)\hat{j}]B_N
     \label{profile-3}
\end{multline}

Here, equation (\ref{eq:bext}) refers to profile-2, and equations (\ref{profile-1}) and (\ref{profile-3}) refer to profile-1 and profile-3, respectively.
Figure \ref{fig:energy_with_distance}(a) shows comparison 
between the $\%$ energy absorption for the three distinct cases of external magnetic field profiles. It can be observed that the absorption efficiency is highest for profile-1 and absorption efficiency decreases for profile-2 and 3. The kinetic energy acquired by the electron species for all three cases at two different times has also shown in other subplots of this figures \ref{fig:energy_with_distance} $(b,c)$. These figures also illustrates that the cut-off energy of the electrons is higher when the distance between vacuum plasma interface and resonance  is less. This can be understood by the variation in refractive index $n_{RX}^2$ for each case as shown in figure \ref{fig:ref_index_shift} (b). By changing the profile, the resonance layer also shifts from its previous location. This implies that if the resonance layer is closer to the point where the EM wave enters the plasma, there will be a longer region over which the EM wave can interact with the resonance and thus a larger fraction of energy will be transferred to the electrons. Also, at the vacuum-plasma interface refractive index $n_{RX}^2$ dictates the fraction of incident EMF energy converted to the transmitted slow wave, as depicted in figure \ref{fig:ref_index_shift} (a). Thus, maximum absorption would be achieved when the resonance layer is closer to the vacuum plasma interface.

 \begin{figure*}
  \centering
  \includegraphics[width=12cm]{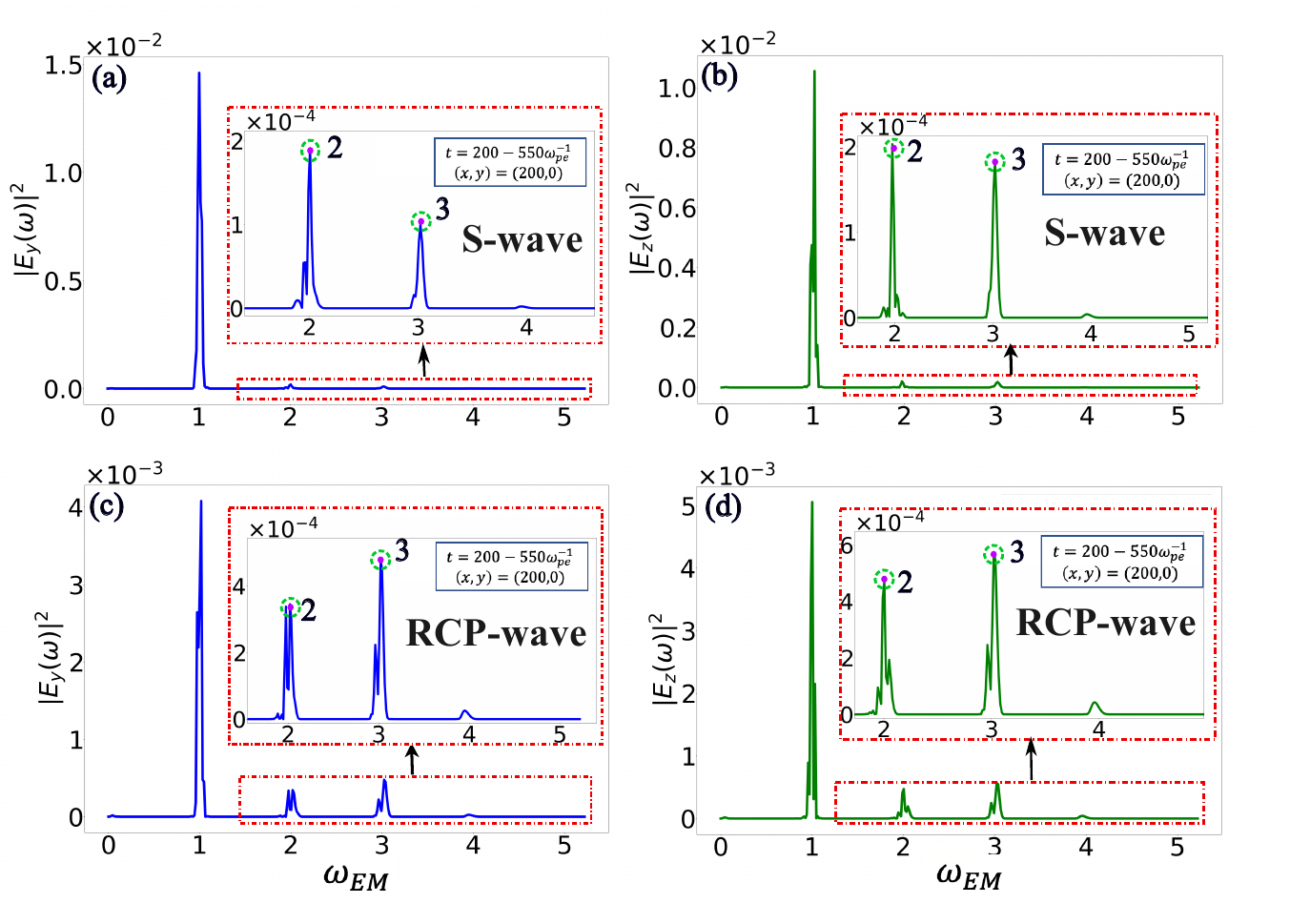} % Images in 100% size
  \caption{Figures $(a,b)$ depict the fast Fourier transform of $E_y$ and $E_z$ components of  the S-polarized wave from $200$ to $550\omega_{pe}^{-1}$ at  $x=200, y=0 $. Similarly, figures $(c,d)$ depict the fast Fourier transform  of the $E_y$ and $E_z$ components of the RCP wave from $200$ to $550\omega_{pe}^{-1}$ at $x=200, y=0 $.}
\label{fig:3rd_harmonic_s-wave}
\end{figure*}

\begin{figure*}
  \centering
  \includegraphics[width=12cm]{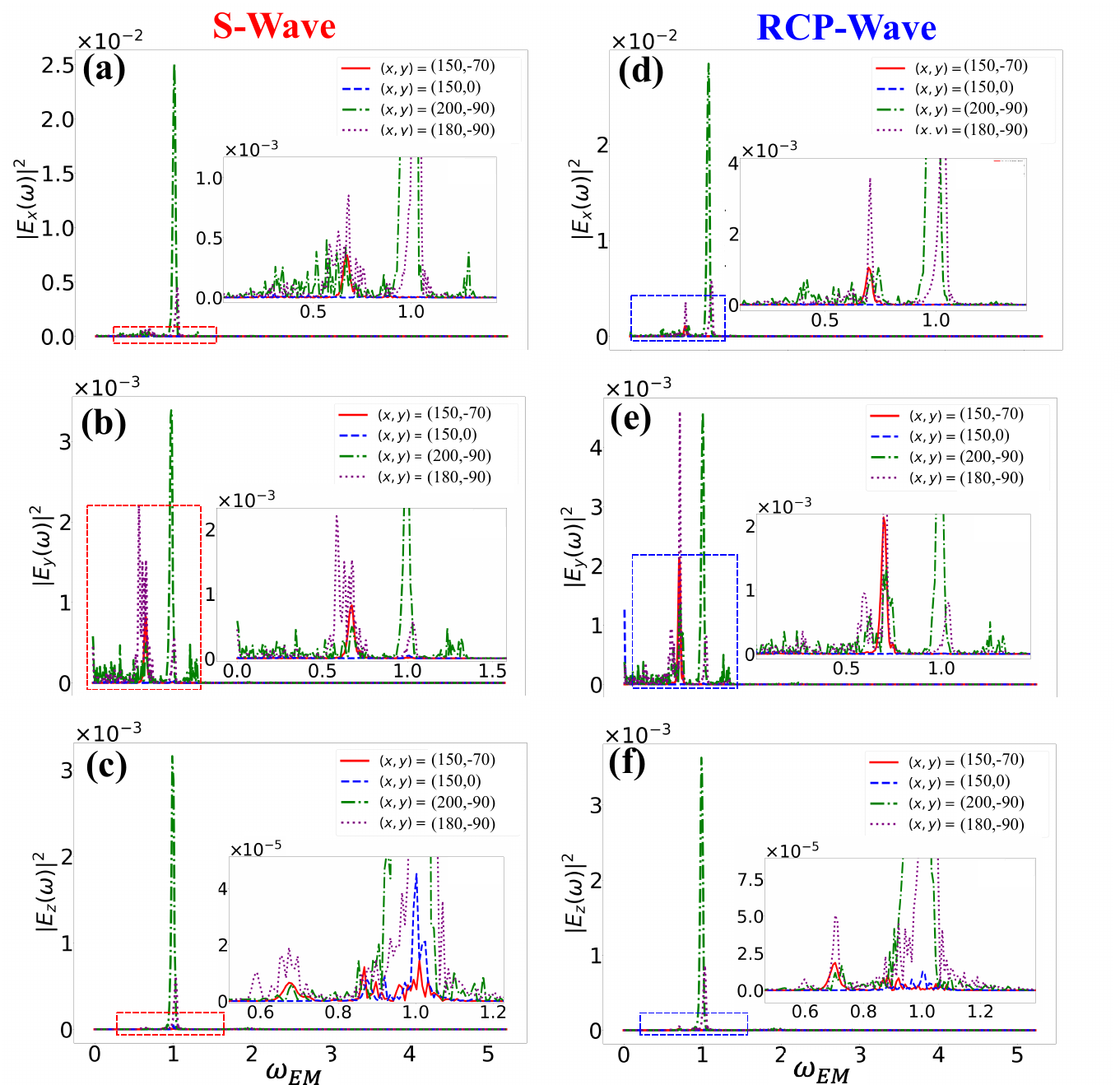}%Images in 100% size
  \caption{ Time fft of the absorbed S-polarized wave $(a,b,c)$ and RCP-wave $(d,e,f)$ near resonance layer at multiple locations for the time window of $700$ to $1400\omega_{pe}^{-1}$.}
\label{fig:time_fft_s-wave}
\end{figure*}

\begin{figure*}
  \centering
  \includegraphics[width=12cm]{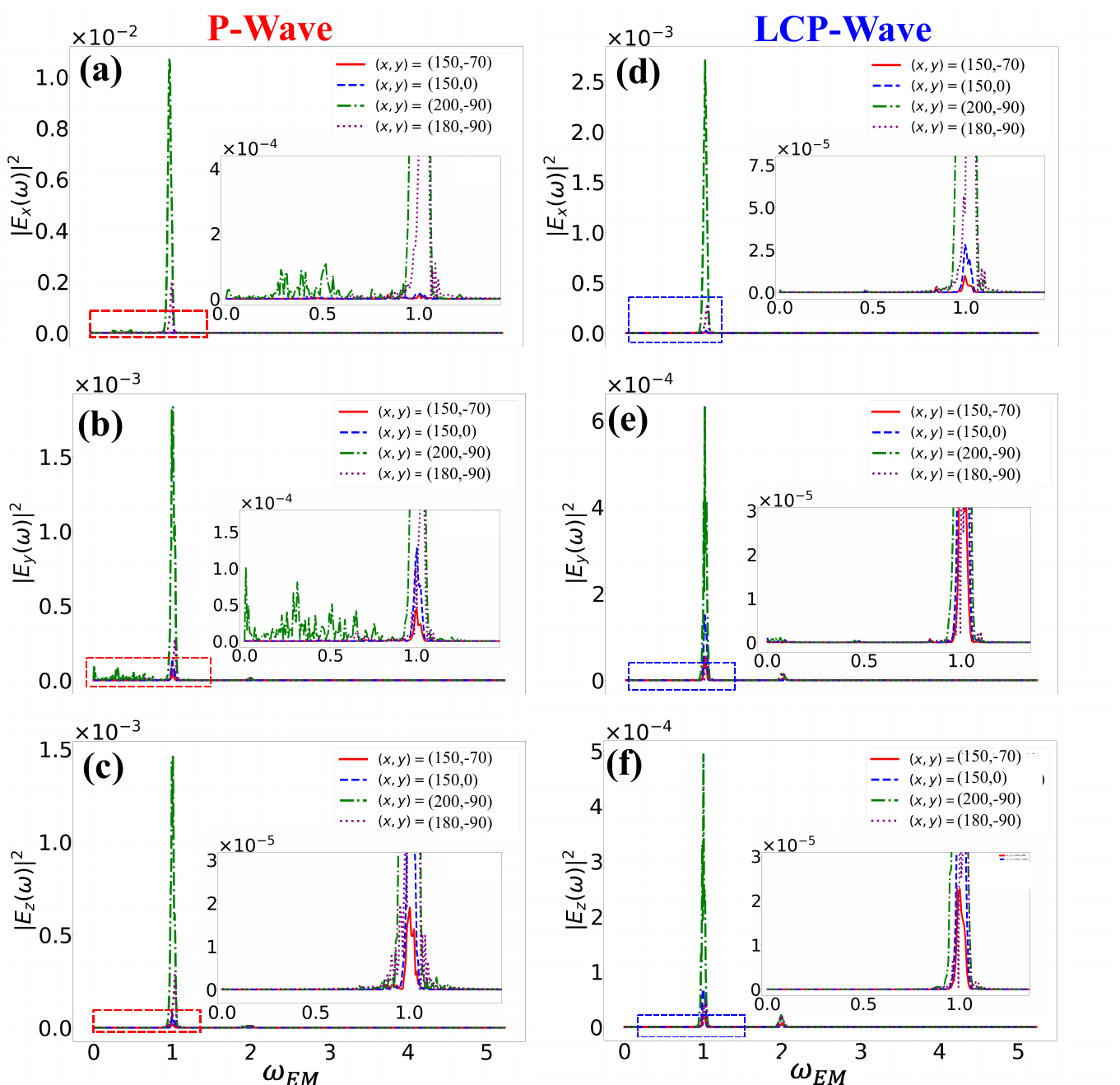}%Images in 100% size
  \caption{ Time fft of the  P-polarized wave $(a,b,c)$ and LCP-wave $(d,e,f)$ near resonance layer at multiple locations for the time window of $700$ to $1400\omega_{pe}^{-1}$.}
\label{fig:time_fft_p-wave}
\end{figure*}
  
\subsection{Nonlinear processes }\label{sec:nonlinear-processes}
 At high intensity EM waves, nonlinear processes may also occur. Furthermore, the choice of an inhomogeneous magnetic field makes the gyrofrequency a function of position. This can trigger the phase mixing processes inside the plasma, making the spectrum broader. We identify some of these processes in this section. We focus on the RCP and the S-polarized EM waves, which have shown efficient absorption of the EM wave energy by the plasma due to the presence of an oblique resonance layer in our study.  

 The nonlinearity of the plasma medium will generate additional frequencies in the system. To observe the generation of other frequencies, we have evaluated Fourier transform of the electric field signal observed inside the plasma medium. We specifically chose the time interval where the absorption was high. 
 Thus, the time interval of  $(t=200-600 \omega_{pe}^{-1})$ and   $(t=700-1400 \omega_{pe}^{-1})$ was chosen to collect the signal, which was then  Fourier analyzed to ascertain the frequency spectrum. 

 For figure \ref{fig:3rd_harmonic_s-wave} the data was collected from the spatial location of $x=200, y=0 $. It clearly shows that both the second and the third harmonic have been excited near the vacuum-plasma interface. Since these higher harmonics have frequencies much greater then gyro-frequency $(2\omega_{EM},3\omega_{EM}>\omega_{ce})$, they do not follow the magnetized plasma dispersion curve and propagate undeflected, exiting from the right boundary.  A detailed study of the higher harmonics generation at the vacuum plasma interface in magnetized plasma has been carried out earlier in \cite{maity2021harmonic, dhalia2023harmonic}. Our simulations also support this. It is thus clear that at a higher intensity (when the nonlinear effects will be more pronounced), a significant portion of the energy will get lost through the process of harmonic generation. This is illustrated in the next subsection. 

The simulation data also identifies the occurrence of the phenomena of phase mixing \cite{maity2012breaking}. The frequency spectra of $|E_x(\omega)|^2$, $|E_y(\omega)|^2$, and $|E_z(\omega)|^2$ were analyzed at four distinct spacetime locations and are shown in figure (\ref{fig:time_fft_s-wave}). The fourier analysis has been carried out for the time interval $t=700$ to $1400 \omega_{pe}^{-1}$ for both RCP waves and S-polarized waves. It is clear from the figure that the frequency spectrum gets broadened with time at a later stage. %We feel this occurs due to phase mixing processes.
We observe that at all points near resonance layer fundamental frequency $\omega_{EM}$ peaks, but on the other hand, we observe that a downshifted low-frequency wave at one distinct peak near $\sim 0.67\omega_{EM}$ is present in the frequency spectra of $|E_x(\omega)|^2$, $|E_y(\omega)|^2$ and a small hump near up-shifted high-frequency wave  $\sim 1.3\omega_{EM}$ is present at location $(200,-90)$. This could be a signature of parametric decay instability (PDI), which generates near the upper hybrid resonance \cite{hansen2017parametric,senstius2019particle}. However, in our geometry, the resonance layer need not to be represent a upper hybrid layer. Thus, it is unclear whether the PDI or inhomogeneity of $B_{ext}$, or both  are responsible for breaking of the electrostatic wave.  We have also plotted time FFT's for P-polarized and LCP wave in figure \ref{fig:time_fft_p-wave}. Since both the P-polarized wave and LCP wave generate significantly less electrostatic generation, as shown in Figure \ref{fig:dive}, we do not observe broadening of the frequency spectrum or breaking of electrostatic waves in either case.

\begin{figure*}
  \centering
  \includegraphics[width=12cm]{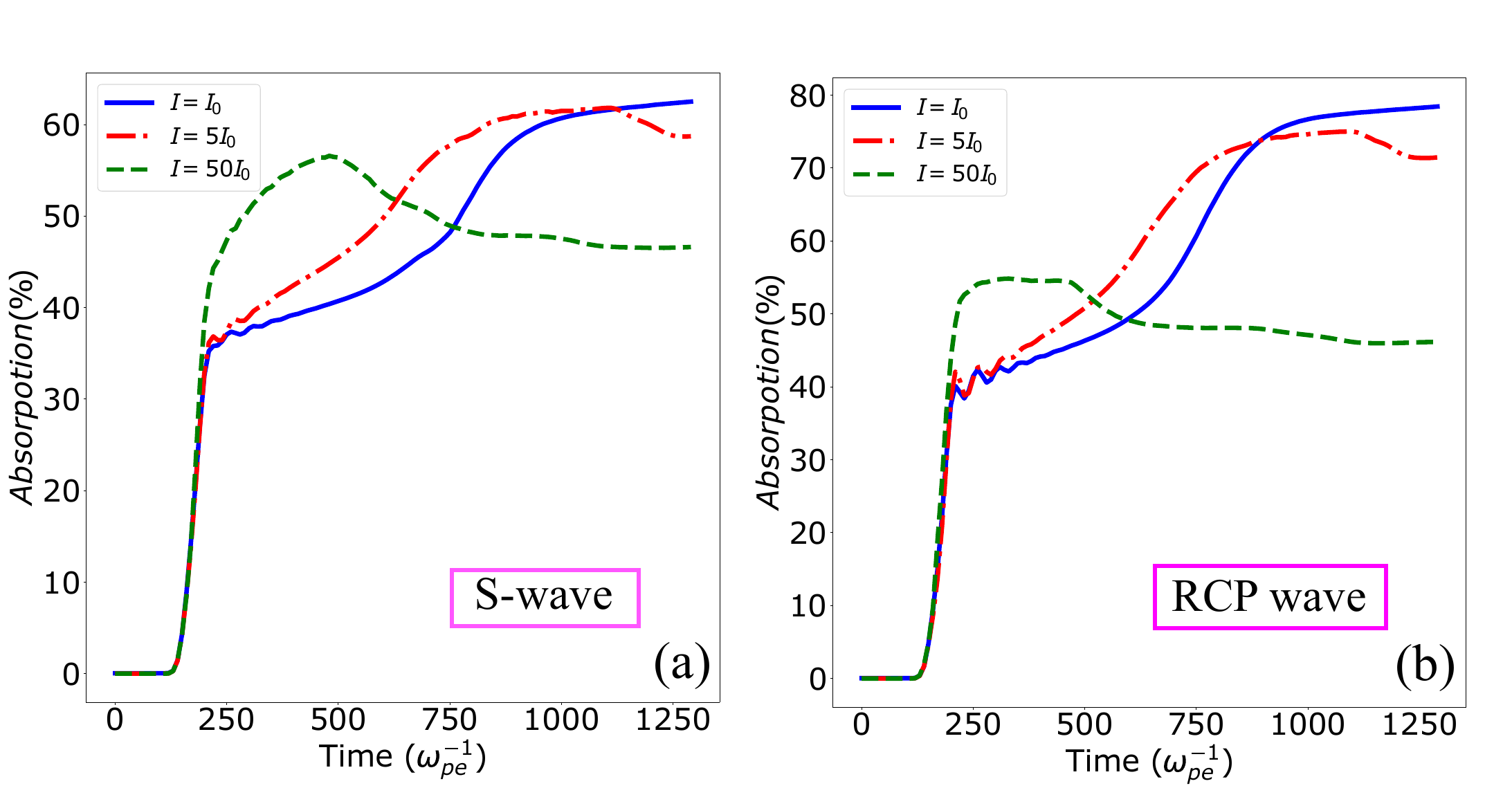}% Images in 100% size
  \caption{ Absorption $\%$ of an S-polarized and RCP wave has been plotted for three different intensities $(I=I_0,5I_0,50I_0)$ while keeping all other parameters fixed. }
\label{fig:energy_absorption_with_intensity}
\end{figure*}
\begin{figure*}
  \centering
  \includegraphics[width=12cm]{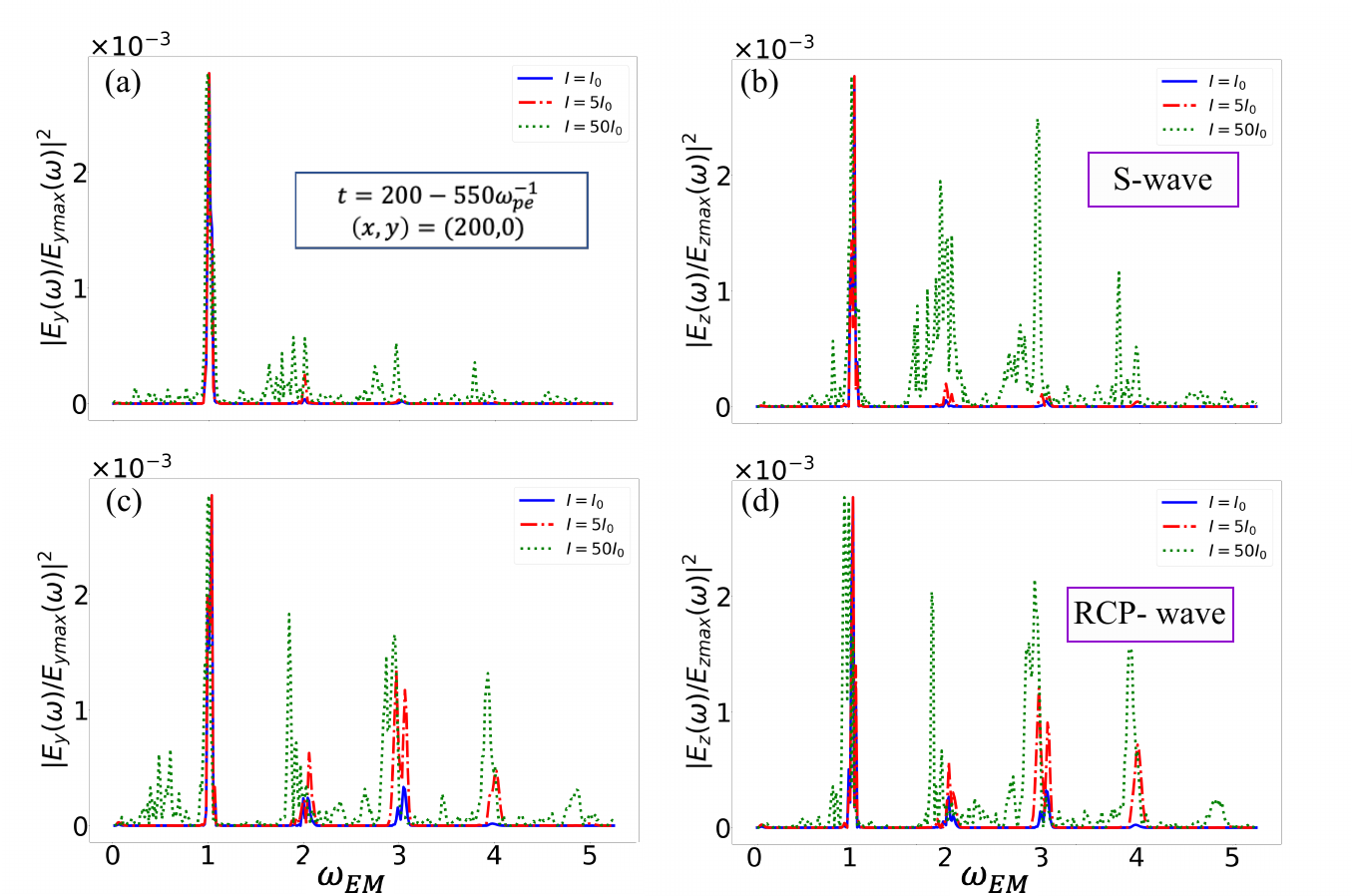}% Images in 100% size
  \caption{Time fft of  $E_y$ and $E_z$ has been plotted for both S and RCP waves at  $x=200, y=0$ for the intensities $(I=I_0,5I_0,50I_0$). }
\label{fig:time_fft_intensity}
\end{figure*}

\subsection{Energy absorption at higher intensity }\label{sec:absorption_with_intensity}
When the intensity of the incident EM wave pulse is increased, though the available energy is high the nonlinear processes described in subsection (\ref{sec:nonlinear-processes}) would become dominant. A significant portion of the energy would then get transferred into the higher harmonics. This would simply result in the EM wave leaking out from the boundary.   We carried out some studies with different intensities of the incident pulse. We observe that the energy gets converted predominantly to higher harmonics at higher intensity of the incident EM wave, thereby reducing the efficiency of the energy absorption. 

 In Figure \ref{fig:energy_absorption_with_intensity} $(a)$ and $(b)$, the temporal evolution of the percentage of energy absorbed by the electrons in the plasma is illustrated for three different intensities $(I=I_0, 5I_0, 50I_0)$ of S-polarized and RCP waves, respectively. There is a notable decrease in the percentage of energy absorption with increasing intensity for both S-polarized and RCP waves. For the S-polarized EM wave, the absorption decreases from $62.1\%$, to $59.9\%$, and further to $46.5\%$. Similarly, for the RCP wave, absorption decreases from $78\%$, to $72.2\%$, and further to $46\%$. At $50I_0$, absorption becomes almost the same for both RCP and S-polarized waves. The reason behind this is that, with increasing intensity, surface phenomena become more active, leading to increased efficiency of the higher harmonics \cite{dhalia2023harmonic}. Since higher harmonics generated at the interface do not contribute to energy transfer and pass through the plasma unhindered, less energy gets  
absorbed by the bulk plasma. The increased intensity of harmonic generation can be observed from the plots of time FFT's of $|E_y(\omega)/E_{ymax}(\omega)|^2$ and $|E_z(\omega)/E_{zmax}(\omega)|^2$ in figures \ref{fig:time_fft_intensity} $(a)$, $(b)$, $(c)$, and $(d)$. Here, $E_{ymax}(\omega)$ and $E_{zmax}(\omega)$ are the amplitudes of the incident EM wave at initial time $t=0 \omega_{pe}^{-1}$ in vacuum.  

\section{Summary}\label{sec:summary}
In this paper, a comprehensive study of EM waves interactions with magnetized plasmas in the presence of an inhomogeneous external magnetic field has been conducted.  Two-dimensional PIC simulations using the OSIRIS 4.0 platform have been carried out for this purpose. Our analysis reveals that the presence of magnetic field inhomogeneity can enhance the bulk energy absorption process when the wave pulse is appropriately polarized. Both the theoretical and the simulation analysis are in agreement. This has been illustrated using the presence of an oblique resonance layer for the slow wave excited at the plasma surface. The efficiency of absorption can be improved by shifting the resonance layer closer to the plasma surface from which EM wave enters in plasma. At higher intensities of the incident wave, nonlinear processes such as harmonic generation continuously drain the energy. It is shown that the efficiency of the absorption process reduces with the increased EM wave intensity as most of the energy gets converted into higher harmonics at the plasma vacuum interface. As the higher harmonics observe no resonance layer and thus the energy simply leaks out with the wave as they propagates out of the plasma from the other end. 

%Study has been made with different cases of polarization of Electromagnetic wave pulse. Maximum energy absorption in electrons has been found to be for Right cirularly polarized and S-polarized EM wave of nearly $80\%$ and $65\%$ respectively. For P-polarized and left circularly polarized EM wave pulse absoprtion was merely $22\%$ and $7\%$ has been found. At the vacuum-plasma surface generation of higher-harmonics of order $2^{nd},3^{rd}$ and $4^{th}$ order has also been detected which are consistent with earlier study by \citep{maity2021harmonic,dhalia2023harmonic} for magnetized plasma. We have also showed that later in time near UHR and ECR layers low frequency electrostatic modes also get excites which is responsible for irreversible energy transfer in electrons. Efficiency of energy absorption by shifting the location of resonance layers for a s-polarized EM wave can be controlled in plasma. It has been established that increasing the intensity of incident EM wave pulse reduces net energy absorption in magnetized plasma because of increasing efficiency of scattered High harmonics electromagnetic radiations. Thus we believe that  our simulations results would prove beneficial for industrial plasma thruster applications and fusion devices. 

 \section*{Acknowledgements}
 \indent The authors would like to acknowledge the OSIRIS Consortium, consisting of UCLA and IST (Lisbon, Portugal), for providing access to the OSIRIS-4.0 framework, which is the work supported by the NSF ACI-1339893. AD would like to acknowledge her  J.C. Bose fellowship grant of AD(JCB/2017/000055) as well as CRG/2022/002782, grant of the Department of Science and Technology (DST) Government of India. The authors would like to thank IIT Delhi HPC facility for computational resources. T.D. also wishes to thank the Council for Scientific and Industrial Research (Grant no- 09/086/(1489)/2021-EMR-I) for funding the research.  We thank the anonymous reviewer for their insightful remarks, which led to the significant improvement of this manuscript.

\section*{Conflict of Interest}
 \indent Authors report no conflict of interest

\bibliographystyle{apsrev4-1}

\bibliography{jpp-instructions.bib}

\end{document}